\documentclass[reprint, amsmath,amssymb, aps, superscriptaddress
]{revtex4-2}

\usepackage[normalem]{ulem}
\usepackage{ragged2e}
\usepackage{multirow}
\usepackage{array}
\usepackage{comment}
\usepackage{graphicx}
\usepackage{amsmath}
\usepackage{amssymb}
\usepackage[none]{hyphenat}
\usepackage{xcolor}
\usepackage{hyperref}  
\usepackage{natbib}
\usepackage{tcolorbox}
\usepackage{enumitem}
\usepackage{float}
\usepackage{tikz}
\usepackage{tikz-3dplot}
\usetikzlibrary{positioning}
\usepgflibrary{shapes.geometric}
\hypersetup{colorlinks=true,urlcolor=blue,citecolor=blue,linkcolor=blue} 

\usepackage{lipsum}
\usepackage{siunitx}
\usepackage{subcaption}
\usepackage[]{caption}
\makeatletter
\long\def\@makecaption#1#2{%
  \vskip\abovecaptionskip
  \begingroup
    \justifying
    \small
    \setlength{\parindent}{0pt}%
    \setlength{\parskip}{0pt}%
    \sbox\@tempboxa{#1. #2}%
    \ifdim\wd\@tempboxa>\hsize
      \noindent #1. #2\par
    \else
      \noindent\hb@xt@\hsize{\box\@tempboxa\hfil}%
    \fi
  \endgroup
  \vskip\belowcaptionskip
}
\makeatother

\begin{document}
\sloppy
\preprint{APS/123-QED}

\title{Examining Student and AI Generated Personalized Analogies in Introductory Physics}


\author{Amogh Sirnoorkar}
\affiliation{Department of Physics and Astronomy, Purdue University, West Lafayette, Indiana - 47907}
\affiliation{Department of Curriculum and Instruction, Purdue University, West Lafayette, Indiana - 47907}

\author{Winter Allen}
\author{Syed Furqan Abbas Hashmi}
\affiliation{Department of Physics and Astronomy, Purdue University, West Lafayette, Indiana - 47907}

\author{N. Sanjay Rebello}
\affiliation{Department of Physics and Astronomy, Purdue University, West Lafayette, Indiana - 47907}
\affiliation{Department of Curriculum and Instruction, Purdue University, West Lafayette, Indiana - 47907}

\date{\today}
\begin{abstract}

Comparing abstract concepts (such as electric circuits) with familiar ideas (plumbing systems) through analogies is central to practice and communication of physics. Contemporary research highlights self-generated analogies to better facilitate students' learning than the taught ones. ``Spontaneous’’ and ``self-generated’’ analogies represent the two ways through which students construct personalized analogies. However, facilitating them, particularly in large enrollment courses remains a challenge, and recent developments in generative artificial intelligence (AI) promise potential to address this issue. In this qualitative study, we analyze around 800 student responses in exploring the extent to which students spontaneously leverage analogies while explaining Morse potential curve in a language suitable for second graders and self-generate analogies in their preferred everyday contexts. We also compare the student-generated spontaneous analogies with AI-generated ones prompted by students. Lastly, we explore the themes associated with students’ perceived ease and difficulty in generating analogies across both cases. Results highlight that unlike AI responses, student-generated spontaneous explanations seldom employ analogies. However, when explicitly asked to explain the behavior of the curve in terms of their everyday contexts, students employ diverse analogical contexts. A combination of disciplinary knowledge, agency to generate customized explanations, and personal attributes tend to influence students’ perceived ease in generating explanations across the two cases. Implications of these results on the potential of AI to facilitate students' personalized analogical reasoning, and the role of analogies in making students notice gaps in their understanding are discussed.       

\end{abstract}

\maketitle














\section{Introduction}
\label{sec:intro}

Making sense of an abstract concept (e.g., electric current) by comparing it to a familiar idea (water flow) is a common practice in physics. Such comparisons facilitate integration of new information into the knowledge system by bridging unknown ideas with familiar ones~\cite{podolefsky2007analogical}. Consequently, comparisons are extensively employed by teachers and informal educators in communicating complex ideas to students and non-disciplinary audiences. Comparisons also form a key component of physicists' practices towards generating new knowledge~\cite{verostek2022making}. Given their extensive use in learning environments, several studies have demonstrated the effectiveness of comparisons in enhancing students’ content understanding in physics~\cite{serbin2024pedagogical,brown1989overcoming}. 

Comparisons can be categorized into similarities and analogies~\cite{gentner2017analogical}.  Similarity statements such as the ``{\em electronic structure of Hydrogen atom is like that of Deuterium}'' map both the object features (such as the number and orbital trajectory of electrons) and the underlying abstract relational features (electrostatic force binding the nucleus and electrons). On the other hand, analogies such as ``{\em Hydrogen atom is like our planetary system}'' predominantly map relational features (centripetal forces) as compared to object features (color of the central, massive entity). Analogies are thus sophisticated comparisons rather than similarity statements, as they establish associations between two dissimilar systems based on the underlying non-trivial, higher order, and abstract relations. Theoretically, an analogy is a mapping between a familiar idea known as the ``Base'' and a relatively unknown idea, the ``Target''. In the above analogy, while the planetary system represents the Base, the atomic model acts as the Target~\cite{gentner1983structure}.  

Studies in physics education have highlighted several benefits and caveats on leveraging analogies for students’ learning. To begin with, analogies can be generative as they aid students in drawing novel inferences about the {\em Target} domain by extending ideas from the {\em Base} domain~\cite{podolefsky2006use}.  This can lead to students' enhanced content understanding and better problem-solving outcomes. However, self-generated analogies tend to better facilitate students' learning than the taught ones. This can be due to various reasons including students' inability to see the analogy's relevance in the given context~\cite{brown1994facilitating} or its limitations in explaining a phenomenon~\cite{glynn1989teaching}. 

However, facilitating students' generation of personalized analogies present challenges (particularly in large enrollment courses) such as attending to individual student's varied lived experiences and consequently their preferred analogical contexts. Recent advancements in Generative Artificial Intelligence (AI) have promised a huge potential in addressing this issue. AI platforms such as ChatGPT are large language models trained on huge datasets which enable them to produce human-like and personalized responses (albeit sometimes incorrect) based on user inputs. Given their capabilities, recent studies in physics education have explored their utility in generating and solving physics assessments~\cite{kortemeyer2023can,sirnoorkar2024student}, or providing personalized feedback~\cite{wan2024exploring,sirnoorkar2025feedback}.  

However, studies focused on comparing student- and AI-generated analogies based on students' preferences in physics education remain scarce. Furthermore, efforts on facilitating students' personalized analogies about abstract college-level physics concepts remain relatively underexplored. Also, research on students' perceived affordances and constraints in generating personalized analogies remains relatively limited.  

We address the aforementioned gaps in the literature by qualitatively investigating how introductory students leverage analogies when asked to explain the Morse potential curve in a language suitable for second graders. We expected that when asked to do so, students would resort to making spontaneous comparisons with second graders' lived experiences to effectively communicate a complex idea to such a non-disciplinary audience. We then contrast student-generated descriptions of these concepts with the AI-generated ones (also generated and reported by students). Furthermore, we analyze students' self-generated analogies, while they described the behavior of the curve in terms of their preferred everyday contexts.   Lastly, we investigate students’ perceived affordances and constraints in coming up with spontaneous and self-generated analogies. 

This manuscript is structured as follows: In the next two sections, we present the background on personalized analogies and articulate our research questions. In Sections~\ref{sec:theory}~and~\ref{sec:design-framework} we detail theoretical and design frameworks guiding our study. We then discuss the study's context along with methods of our data collection and analysis in Section~\ref{sec:methods}. In Section~\ref{sec:results}, we present our results and then discuss them in Section~\ref{sec:discussion}. Lastly, in Section~\ref{sec:conclusions}, we provide the concluding remarks along with the study's implications, limitations, and future work.   

\section{Background}
\label{sec:background}

Analogies establish associations between a familiar domain ({\em Base}) and a relatively unfamiliar one ({\em Target}). Constructing such mappings constitutes a central epistemic practice in physics, as they enable learners to make sense of novel and abstract ideas. Prior research has consistently underscored their role in supporting students’ conceptual understanding and problem-solving in physics~\cite{gentner2014flowing}. However, several challenges accompany the effective use of analogies in learning.

Firstly, students may not always employ analogies as a reasoning tool~\cite{gick1980analogical}, and when they do, there is a risk of drawing unintended inferences~\cite{zook1991effects}. For example, when analogizing the atomic model to a planetary system, students might incorrectly infer that electrons, like planets, attract each other. Brown~\cite{brown1994facilitating} observed that analogies that seem intuitively appropriate to experts may not resonate similarly with students. Thus, learners may not always appreciate the limits or scope of analogies~\cite{glynn1989teaching}.  Brown and Clement~\cite{brown1989overcoming} argue that for productive reasoning with analogies, students must participate in an epistemologically dynamic environment rather than simply accommodating a provided analogy.

To mitigate the aforementioned issues, researchers have advocated for the ``production'' paradigm over the ``reception'' paradigm~\cite{blanchette2000analogies}. That is, instead of introducing an existing analogy and expecting students to interpret and use it in an intended manner (``reception''), students should generate their own analogies by organizing a studied phenomenon and exploring its properties (``production''). Such an approach is beneficial as this tends to be  a primary way disciplinary experts generate analogies~\cite{verostek2022making}. Furthermore, since it was students who generated analogies in the first place, they are more likely to be aware of the exact mapping structure between the two domains, thereby also being aware of its potential limits.  

Several studies have highlighted the benefits of the production paradigm. Coll {\em et al.}~\cite{coll2005role} note that analogies are most effective when students construct and critique their own and scientists’ models.  Haglund and Jeppsson~\cite{haglund2012using} observed that when pre-service teachers generated their own analogies about thermodynamic processes, they took ownership of their learning and engaged in explorative discourse.  Lastly, explaining known ideas by associating them with everyday experiences forms a key component of ``sensemaking'', a valued cognitive process in science education~\cite{defining,sirnoorkar2023theoreticaljournal,sirnoorkar2023sensemaking,methodssirnoorkar}.

Along the same lines, researchers have explored the production paradigm by explicitly focusing on ``spontaneous'' and ``self-generated'' analogies~\cite{haglund2013collaborative}. Spontaneous analogies are the mappings or associations that students generate without explicit instruction. ``Self-generated'' analogies, on the other hand, correspond to the mappings student generate when explicitly asked to bridge studied concepts with lived antecedent experiences. Both types of analogies have been observed to facilitate personalized learning experiences in science learning environments~\cite{clement201323,kind2007creativity}.    


\section{Research Questions}
\label{sec:rqs}

In the rest of this manuscript, we answer the following research questions :

\begin{enumerate}
    \item[\bf RQ1] {\em How do spontaneous analogies employed by introductory students to explain the Morse potential curve in a language suitable for second graders compare to the ones generated by AI in the same context?} 

    \item[\bf RQ2] {\em To what extent do students self-generate analogies when asked to explain the Morse potential curve in terms of their preferred everyday contexts?} 

    \item[\bf RQ3] {\em What are the trends and emergent themes associated with students’ perceived degree of ease  while explaining a complex concept such as Morse potential curve in a language suitable for second graders and in terms of their preferred domains? }
\end{enumerate}

\section{Theoretical Framework}
\label{sec:theory}

\subsection{Structure-Mapping Theory}
\label{subsec:structure-mapping}

Comparisons such as ``{\em atomic structure is like a planetary system}'' establish parallels between two distinct contexts. When establishing such comparisons, one can map both object attributes such as the shape of trajectories of atoms and planets, as well as higher order causal relations or ``relational predicates'' such as the presence of central forces. Comparisons that map both object attributes and relational predicates correspond to similarity statements. However, the ones that predominantly map relational predicates as compared to object attributes are termed as analogies. Structure-Mapping theory~\cite{gentner1983structure} provides a theoretical account of comparisons, particularly analogies.

According to this theory, the difference between literal similarity and analogy is not dichotomous, but rather a continuum. For example, the statement ``{\em My tea kettle is like your tea kettle}'' represents a literal similarity statement where there is a huge overlap between the structural and functional features of both the kettles. However, the comparison ``{\em A steam engine is like a tea kettle}'' highlights a moderate overlap between functioning of a steam engine and a tea kettle. Lastly, the statement ``{\em Anger is like a tea kettle}'' represents an analogical comparison as the focus is more on the higher order casual relational structures (such as anger bursting out upon a threshold) rather than the structural ones.

Thus, an analogy ``{\em A T is like a B}'' defines a mapping from B to T where T is referred to as the ``Target’’ and B as the ``Base''. The {\em Target} corresponds to a relatively unfamiliar idea or domain that is being explicated, and the {\em Base} corresponds to the familiar idea serving as the knowledge source. This mapping between the two domains is guided by the following constraints~\cite{markman2000structure}:
\begin{enumerate}
    \item {\em Relational similarity}. Analogies involve relational commonalities; object commonalities are optional.

    \item {\em Structural consistency}. Analogical mapping involves one-to-one correspondence and parallel connectivity.
    \begin{enumerate}
        \item {\em One-to-one correspondence}. Analogical mapping requires that each element from one domain be matched to at most one element in the other domain. 

        \item {\em Parallel connectivity}. If a pair of attributes or relations is placed in correspondence, then the arguments of those attributes or relations must also be placed in correspondence.
    \end{enumerate}

    \item {\em Systematicity principle}. In interpreting an analogy, connected systems of relations are preferred over sets of isolated relations. This principle highlights a tacit preference for coherence and causal predictive power rather than for sets of surface-level coincidental matches.

    \item {\em Candidate inferences}. Analogical mapping can lead people to draw novel inferences by establishing additional connections between the two representations.
\end{enumerate}

In the current study, we adopt the Structure-Mapping theory in characterizing student- and AI-generated comparisons while explaining the Morse potential curve in a language suitable for second graders and in terms of their preferred contexts.

\subsection{Analogical Retrieval}
\label{subsec:analogical-retrieval}

Analogical retrieval is a cognitive process that elucidates analogy-generation through the following set of overlapping but distinct processes~\cite{forbus1995mac,holyoak1987surface,gentner2017analogical}:

\begin{enumerate}
    \item {\em Retrieval}. Given a current topic in working memory ({\em Target} domain), a person may be reminded of a prior analogous situation ({\em Base}) in the long-term memory.
   
    \item {\em Mapping}. Once the two cases are co-present in working memory, mapping is carried out by aligning the representations. This alignment involves noting the commonalities and differences between the two cases, and projecting inferences wherever necessary from one analog to another.

    \item {\em Evaluation}. Once the mapping is accomplished, the analogy and its inferences are judged for relevance and validity in the target domain.
\end{enumerate}

In the current study, we adopt analogical retrieval in designing the task to facilitate students' analogical reasoning about the Morse potential curve. To facilitate ``{\em Retrieval}'' (the first process), we first asked students to describe the curve in their own words so as to facilitate the ``placing'' of the conceptual idea in their working memory. This task was then followed by asking students to describe the curve to second graders and later in terms of their preferred contexts to further facilitate the ``{\em Mapping}'' phase. This task was then followed by asking them to reflect on the perceived ease of generating explanations across the cases thereby elucidating the ``{\em Evaluation}'' phase. Refer Section~\ref{subsec:data-collection} for details.

\begin{figure}
    \centering
    \includegraphics[scale=0.45]{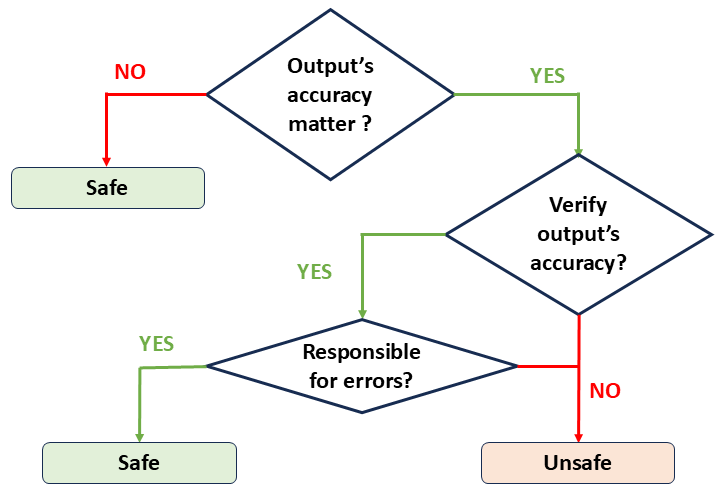}
    \caption{The study's design framework adopted from United Nations Educational, Scientific and Cultural Organization's (UNESCO) guide. The framework has been slightly modified and the original flowchart can be found in~\cite{sabzalieva2023chatgpt}.}
    \label{fig:unesco-framework}
\end{figure}

\section{Design Framework}
\label{sec:design-framework}

Since recent years, generative artificial intelligence (AI) has caught increasing traction in physics education. AI platforms or large language models are trained on a vast amount of data to enable them to generate human-like, customized, sophisticated responses to user queries. Recognizing the accessibility and influence of such powerful tools, the United Nations Educational, Scientific and Cultural Organization (UNESCO) published a quick start guide titled ``ChatGPT and Artificial intelligence in higher Education'' to discuss the potential applications of AI in higher education~\cite{sabzalieva2023chatgpt}. This guide introduces a framework, presented as a flowchart, that outlines the conditions under which AI can be safely and effectively used to enhance student learning. A modified version of the flowchart is highlighted in Figure~\ref{fig:unesco-framework}. The framework is structured around three guiding, question-based criteria that help determine the responsible use of AI in higher education contexts.

The first criterion examines whether the accuracy of the AI’s output is critical in the given context. If accuracy is not essential, AI use is encouraged. Such contexts may include brainstorming creative ideas or prompting AI to simulate student responses while responding to novel classroom activities. When accuracy matter, the second criterion considers whether users have access to sufficient expertise or resources to verify AI-generated outputs. In the absence of such expertise, AI use is discouraged. However, if verification is possible, the final criterion addresses the user’s willingness to assume responsibility for any overlooked inaccuracies. If users are prepared to take responsibility, AI use is deemed appropriate; otherwise, it is not recommended.

We adopt this framework in designing our activity involving students' use of AI in generating descriptions of the Morse potential curve in a language suitable for second graders. Details of the interplay of this framework and the task design are detailed in the next section.

\section{Methods}
\label{sec:methods}

\begin{figure}
     \centering
     \includegraphics[scale=0.45]{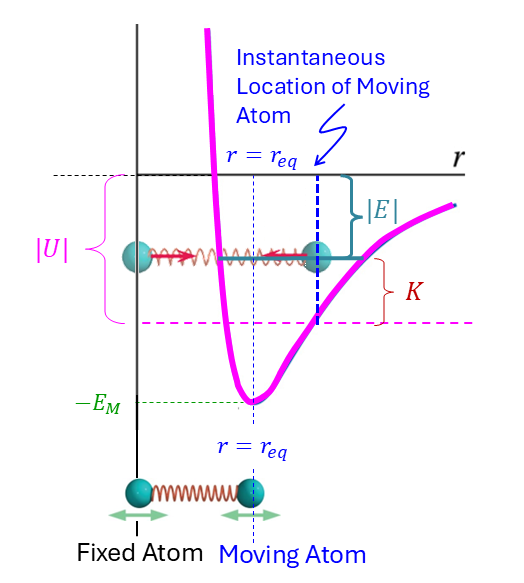}
     \caption{The diagram of Morse potential curve provided to students during data collection. }
     \label{fig:morse-curve}
\end{figure}

\subsection{Context: Morse Potential Curve}
\label{subsec:morse-curve}
The Morse potential energy curve, illustrated in Figure~\ref{fig:morse-curve}, models the interaction between two neutral atoms in a diatomic molecule as a function of their internuclear separation $(r)$. The potential energy $(U(r)$, shown by the solid magenta curve) captures both the short-range repulsive interaction arising from overlapping electron clouds and the longer-range attractive interaction due to van der Waals or bonding forces. The curve reaches a minimum at the equilibrium distance $(r_{eq})$, where the attractive and repulsive forces balance. The corresponding energy value, $(-E_{M})$, represents the bond dissociation energy, i.e., the amount of energy required to separate the two atoms completely. At this equilibrium point, the total force on each atom is zero, and the system is in its most stable configuration.

The figure also depicts how a moving atom oscillates about this equilibrium position. To remain consistent in how this curve was discussed in classroom lectures, a spring illustration was included and retained in the figure. Its total energy $(E$, shown as a  teal horizontal line) remains constant, while its potential energy and kinetic energy continuously exchange as it vibrates. When the atom is displaced from $(r_{eq})$, the increase in potential energy reduces its kinetic energy, and vice versa. The energy difference between the total energy $(E)$ and the potential at that point $\lvert U \lvert$ at any point gives the instantaneous kinetic energy $(K)$ of the atom. 

\begin{figure}
     \centering
     \includegraphics[scale=0.4]{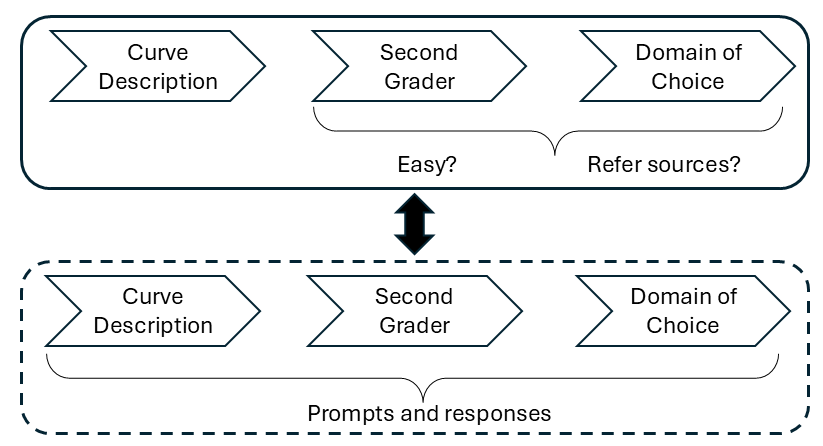}
     \caption{Details of our data. The elements within the solid line indicate the data from students whereas those within dotted lines correspond to the data which students generated from Generative-AI. The double headed arrow signifies students' comparison between each corresponding elements. The data also includes students' perceptions of the usefulness of the activity which has not been highlighted in the above figure. }
     \label{fig:data}
\end{figure}

\subsection{Data collection}
\label{subsec:data-collection}
The data for this study was collected through an online extra credit activity (worth 1\% of the course grade) in a first-semester calculus-based large-enrollment physics course primarily for future engineers, at a large U.S. Midwestern land grant university. The annual enrollment at the time of the study was approximately 2500 students (approximately 1100 in fall and around 1400 in spring). The enrollment of the course in the Spring 2024 semester in which the data were collected is N = 1446. 

The course follows Chabay and Sherwood's \textit{Matter \& Interactions}~\cite{chabay2015matter} and its weekly structure consists of three components: lectures (two 50-minute-long sessions), lab (one 110-minute-long session), and recitations (one 50-minute-long session). Student responses were collected via Qualtrics~\footnote{https://www.qualtrics.com/} on Week 11 of the semester, at which stage, the Morse potential curve was fully introduced. Many students had their first exposure to the curve in this course, as it is not traditionally covered in high school physics curricula. Since students often found this concept challenging on assessments, it was deliberately selected (in consultation with the course coordinator) to support their conceptual understanding. The survey was designed to help students make sense of the curve through both spontaneous and self-generated analogies, supplemented by the use of Generative AI. Students were expected to complete the survey within 60 minutes.

The activity, summarized in Figure~\ref{fig:data}, consisted of 14 open ended questions. Students were provided with the diagram of the curve and were first asked to explain their understanding of the curve including the physical meaning of ``$r_{eq}$'' and its minimum within 150 words. The aim of this task was to not only get a baseline of student understanding of the curve, but to also facilitate them in the first phase of the analogical retrieval process (Section~\ref{subsec:analogical-retrieval}). Students were then asked to describe the curve as they would to a second grader and then in terms of their everyday contexts. These two tasks were aimed at facilitating students in generating spontaneous and self-generated analogies by engaging them in the second phase of the analogical retrieval process (``Mapping''). Subsequently, students were asked to report their perceived ease in generating explanations across the two cases (second grader and domain of choice) and provide appropriate reasoning. This ensured, the third phase ``Evaluation'' of the analogical retrieval process. Information about their references to resources (e.g., textbook, notes, etc.) while generating explanations were also collected.

\renewcommand{\arraystretch}{1.15}
\begin{table}[tb]
\begin{ruledtabular}
\caption{Cohen's $\kappa$ and percentage agreement values (highlighted by ``\%'') for various qualitative codes associated with even- and odd-numbered (50) student responses coded independently by two researchers.}
\label{tab:irr}
\begin{tabular}{
  >{\raggedright\arraybackslash}p{0.55\linewidth}
  @{\hspace{0.5em}} S S
  @{\hspace{0.5em}} S S
}
\multicolumn{1}{c}{} &
\multicolumn{2}{c}{\shortstack{\textbf{Second}\\ \textbf{Grader}}} &
\multicolumn{2}{c}{\shortstack{\textbf{Domain of}\\ \textbf{choice}}} \\

\colrule
\multicolumn{1}{c}{} & {Even} & {Odd} & {Even} & {Odd} \\
\colrule

Presence of comparisons                 & 0.95 & 0.94 & 1.00 & 0.66 \\

Characterization of similarity/analogy           & 0.95 & 0.92 & 1.00 & 1.00 \\

Characterization of perceived ease (Easy/difficult)  & 1.00 & 0.94 & 1.00 & 1.00 \\

Identification of analogical context (\%)  & 0.96 & 1.00 & 1.00 & 1.00 \\

Themes of perceived ease (\%)  & 0.85 & 0.91 & 0.92 & 1.00 \\
\end{tabular}
\end{ruledtabular}
\end{table}

\begin{figure*}[ht]
  \centering
  \subfloat[]{%
    \includegraphics[scale=0.45]{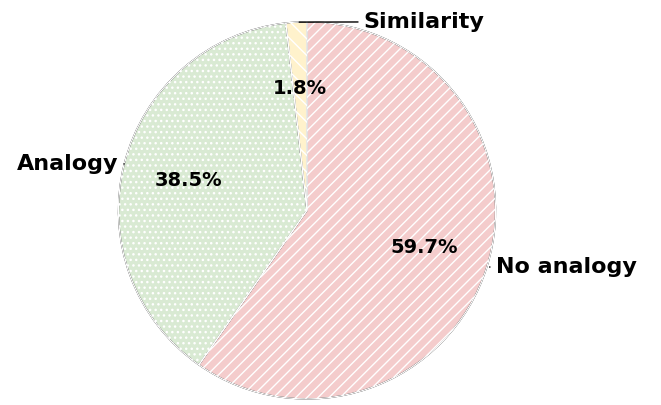}\label{fig:q4-piechart}}
  \subfloat[]{%
    \includegraphics[scale=0.42]{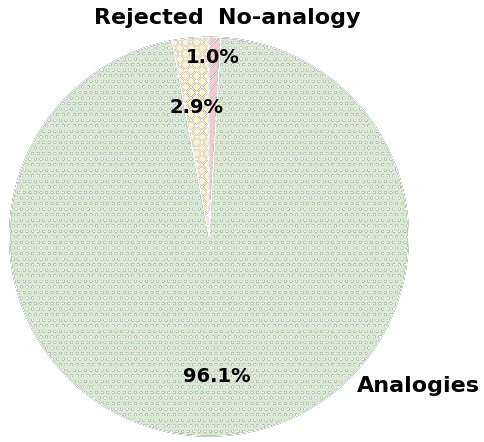}\label{fig:q4-AI-pie-chart}}
  \hfill
  \subfloat[]{%
    \includegraphics[scale=0.42]{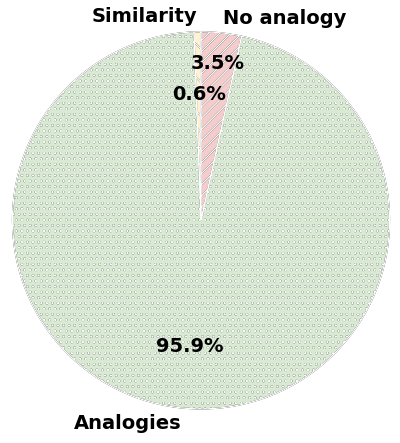}\label{fig:q7-pie-chart}}
  \caption{\justifying%
  Pie charts highlighting the extent of comparisons observed in (a) students' explanations to second graders, (b) AI-generated explanations to second graders, (c) students' explanations in their preferred contexts.}
  \label{fig:three-wide}
\end{figure*}

The second half of the survey asked students to leverage AI to complete the same three activities: (i) describing the curve, (ii) explaining it to a second grader, and (iii) describing it in terms of their preferred everyday contexts. Students’ prompts to the AI and the corresponding AI-generated responses were collected. For each of the three activities, student- and AI-generated responses were presented side by side, and students were asked to identify three qualitative differences between the paired responses. The final task of the survey invited students to reflect on the overall usefulness of the activity for their understanding of the curve and to provide reasons supporting their evaluation.

The AI-focused tasks were intentionally designed to align with the three criteria of the design framework described in Section~\ref{sec:design-framework}. The first criterion was addressed since students were only required to submit AI responses (along with their prompts) and were awarded credit regardless of correctness. In instances where students wished to verify the accuracy of the AI’s output, they had access to relevant course materials, thus addressing the second criterion. Finally, since students engaged in critiquing and reflecting on the AI-generated content they reported, the third criterion was also satisfied. 


\subsection{Data analysis}
\label{data-analysis}
The survey data was then downloaded into a spreadsheet and incomplete or ``blank'' responses were removed. Of the 1243 students, 404 did not respond to all survey questions and were thus removed, leaving us with a total of 839 responses.  To answer the first research question, the first author went through each response generated by students and AI, explaining the Morse potential curve in a language suitable for second graders. The responses were qualitatively coded in terms of whether the description corresponded to a similarity, an analogy, or neither. Statements which demonstrated only literal similarity such as the ``{\em the first half of the curve looks like a slide}'' were marked as similarity statements. If descriptions contained a combination of similarity statements and analogies, they were coded as ``analogy'' and the involved context was noted. When multiple contexts were present in the same description, the predominant context was coded as the representative analogical context. If the response was neither a similarity nor an analogy, but rather reflected only simplification of technical vocabulary, the description was coded as ``Neither''.  A similar process was then repeated to answer the second research question by going through student-generated responses explaining the curve in terms of their preferred everyday contexts. Despite having access to AI-generated analogies about students' preferred contexts, we do not analyze them in this study, as we asked students to explicitly include their preferred contexts  when prompting to AI models.  

To answer the third research question, ``emergent coding’’~\cite{creswell2016qualitative} was adopted to analyze student responses about the perceived ease and difficulty across the two data sets (second graders and domain of choice). The first author initially categorized each response as ``Easy'' or ``Difficult’’ based on the accompanying description. Responses that reflected a combination of both were coded as ``Intermediate’’. Subsequently, responses within each category (Easy, Difficult, and Intermediate) were analyzed to identify emergent codes highlighting the reasons for students’ perceived ease or difficulty. Codes with overlapping meanings were then merged to form a set of mutually exclusive codes, representing distinct themes that explained students’ perceptions across the three categories. Using this finalized codebook, each student’s response for both scenarios (second grader and domain of choice) was recoded to determine the prevalence of these themes in the dataset.

The second and third authors each randomly selected 50 student responses and analyzed them to: (i) determine whether the responses included comparisons, (ii) if so, classify them as similarities or analogies, (iii) identify the analogical contexts when analogies were present, (iv) categorize students’ reflections as indicating ease, difficulty, or an intermediate level, and (v) identify themes associated with their perceived ease. While the second author coded 50 random even numbered responses, the third author coded 50 odd-numbered responses to avoid the overlap. The values of Cohen’s $\kappa$ and percentage agreement for the each of the two sets are summarized in Table~\ref{tab:irr}.


\renewcommand{\arraystretch}{1.1}
 \begin{table*}[tb]
\begin{ruledtabular}
\caption{Exemplar quotes highlighting student explanations without comparisons, entailing similarity, and analogy while explaining Morse potential curve in a language suitable for second grader.}\label{tab:second-grader-exemplar-quotes}

\begin{tabular}{p{0.1\linewidth} p{0.87\linewidth}}

Category & Exemplar Quote  \\
\hline 

No comparisons & ``{\em  The two atoms have the same total energy no matter. The atoms have energy in two forms: from how fast they are moving, and from how far apart they are. The two atoms want to be closer together. But, when they are too far apart from each other, they don't care to be close together anymore. $R_{eq}$ is the distance where two atoms are as far apart from each other as possible while still wanting to be close together. Any farther apart, and they start to care less and less about being together. How much they want to be together can be seen by the purple line. The bottom tip of the purple line is the maximum amount the two atoms feel like they want to be together. They feel this when they are the distance of $r_{eq}$ apart.}'' \\

Similarity & ``{\em  On a playground, I want you to imagine a large slide. Have you seen how it goes down and then up again? This curve is just like the slide on the playground. When you go down the slide, at some point you stop at the bottom. You are not sliding anymore. This curve is very similar. In this case, the potential energy is at a minimum. It is not ``sliding anymore''.}''  \\

Analogy & ``{\em  The sweet spot between these two atoms and how much they want to be together is that distance r(eq). It's like a cat warming itself up at a fireplace, or campfire: too close, and it's too hot and burning itself! Too far away, and it doesn't feel any warmer than if there was no campfire! The minimum, where the curve is at its lowest, represents the energy (in negative magnitude) needed to move the cat from the sweet spot. It takes a lot of energy to pull a cat away from a nice warm spot, especially compared to if it was burning or if it was cold!}''  \\
\end{tabular}
\end{ruledtabular}
\end{table*}

\section{Results}
\label{sec:results}

In this section, we present the results addressing the three research questions outlined in Section~\ref{sec:rqs}. The first question compares student- and AI-generated explanations of the Morse potential curve in language suitable for second graders. Its primary aim is to examine the extent to which students employ spontaneous analogies when explaining a complex concept to an audience with no prior background knowledge, and to what extent AI can support this process. The second research question investigates the extent to which students self-generate analogies when explicitly prompted to explain the curve in contexts of their choice. The goal here is to explore whether students exhibit preferences for particular analogical contexts or express their ideas creatively when provided the agency in selecting them. The third research question examines trends and themes related to students’ perceived ease or difficulty in generating explanations or analogies across the two cases. In each of the following three subsections, we first summarize the overarching results before discussing detailed trends, including exemplar analogies analyzed through the lens of Structure-Mapping Theory, discussed in Section~\ref{subsec:structure-mapping}.  

\subsection{RQ1: Student- and AI-generated curve descriptions to second graders}
\subsubsection{Student descriptions of the curve}
\label{subsec:q4-student}

Broadly, students did not frequently employ comparisons in general, or spontaneous analogies in particular, when explaining the curve to second graders. Among the analogies that were generated, the most common context corresponded to the one depicted in the provided image of the curve (Figure~\ref{fig:morse-curve}).

Of the 839 student responses, 501 (60\%) did not include any comparisons. These explanations primarily simplified technical vocabulary or provided descriptive accounts of the curve. Among the remaining responses, 338 included comparisons, 323 of which were analogies and 15 were similarity statements. Furthermore, within the 323 analogies, 37 featured multiple analogical contexts embedded within a single description. Figure~\ref{fig:q4-piechart} summarizes these trends, while Table~\ref{tab:second-grader-exemplar-quotes} provides exemplar student descriptions across the three categories.    

\begin{figure}
    \centering
    \includegraphics[scale=0.28]{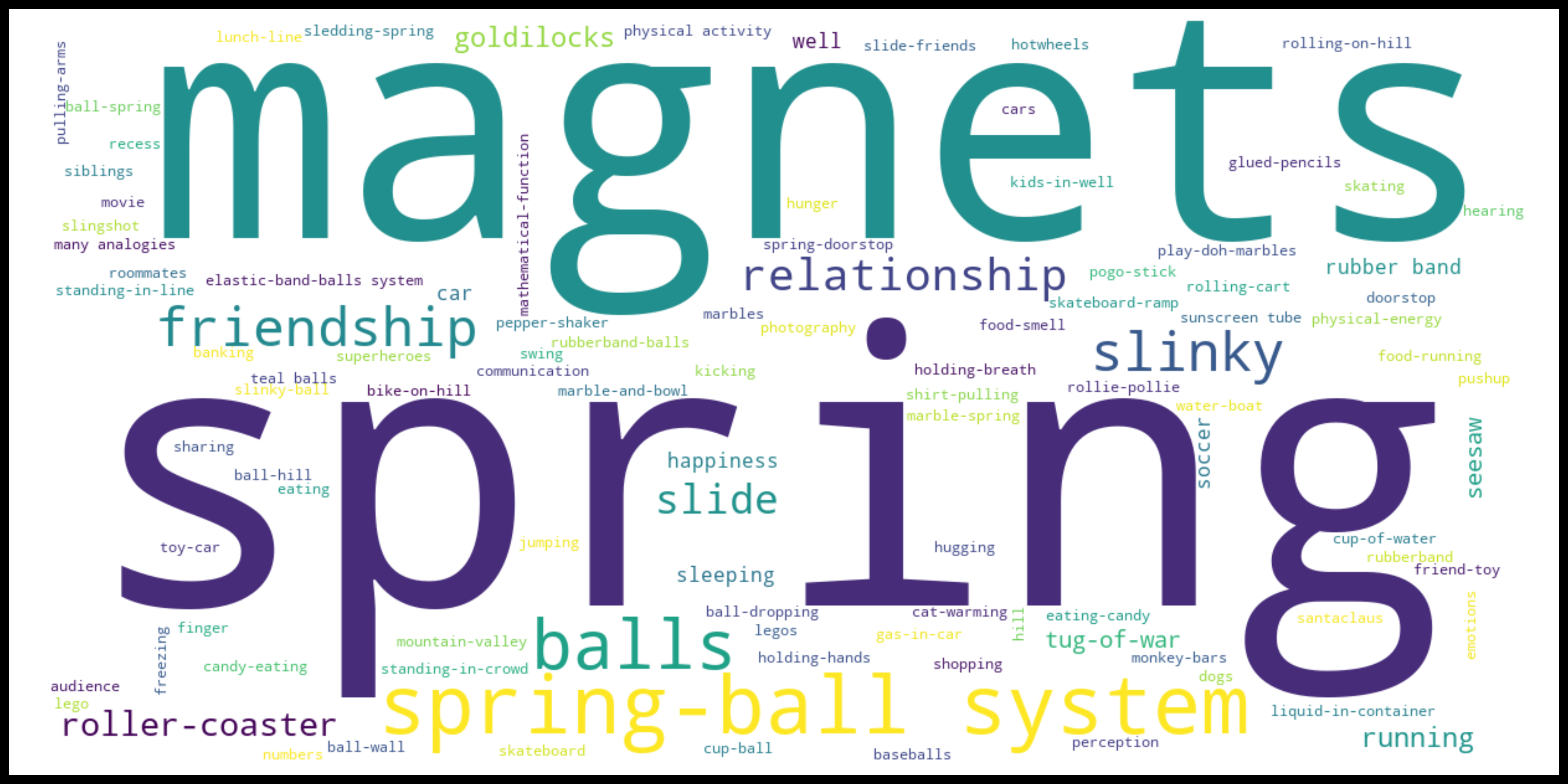}
    \caption{Word cloud representing the various analogical contexts employed by students when describing the Morse potential curve in a language suitable for second graders.}
    \label{fig:q-4-students-wordcloud}
\end{figure}

Figure~\ref{fig:q-4-students-wordcloud} presents the word cloud highlighting the diversity of contexts (identified through manual qualitative coding) in student generated analogies. Evidently, the context of springs (including spring-balls systems) and magnets were the dominant contexts. The frequent use of the spring context likely stems from its depiction in the provided curve during the activity.  Other common analogical contexts included slinky and relationships (including friendship). Below is a student-generated analogy of the curve in the contexts of friendship and spring:
\begin{quote}
    ``{\em When the pink line hits the black line named r, this means that the spring is totally satisfied where its at. It's like if you were hanging out with your friend and were completely happy with your relationship. Left of this point is when the spring wants to extend. This is comparable to when your friend is around you all the time that you get annoyed by them. To the right of this curve is where the spring wants to contract. This would be like if you haven't seen your friend in so long that you want to be with them really badly. However, there is a point where too much distance leads to fading away. If you haven't seen your friend in years, you are not going to be as friendly as you would if it had been only months.}''
\end{quote}

\renewcommand{\arraystretch}{1.1}
 \begin{table*}[tb]
\begin{ruledtabular}
\caption{Exemplar AI responses highlighting the curve descriptions without comparison and through an analogy while explaining Morse potential curve in a language suitable for second grader. There were no similarity comparisons observed.}\label{tab:AI-exemplar-responses}

\begin{tabular}{p{0.1\linewidth} p{0.87\linewidth}}

Category & Exemplar Quote  \\
\hline 

No comparisons & ``{\em The Morse potential energy graph depicts the relationship between potential energy (y-axis) and atomic separation distance (x-axis) for diatomic molecules. It exhibits a characteristic curve with a well depth, representing the minimum potential energy when the atoms are at equilibrium distance ($r_{eq}$). The curve rises steeply at short distances due to repulsion between the atoms, reaches a maximum at the equilibrium distance, and then gradually levels off as the atoms move further apart due to weak van der Waals forces. The equilibrium distance ($r_{eq}$) is where the attractive and repulsive forces balance, corresponding to the bond length at which the molecule is most stable. This distance is a crucial parameter in understanding molecular bonding and predicting molecular properties.}'' \\

Analogy & ``{\em  Alright, imagine two atoms holding hands, but they're not too close or too far apart; they're just right, like Goldilocks finding the perfect bed. The energy diagram shows how much energy they have when they're at different distances from each other. The bottom of the curve is like the coziest spot for these atoms – it's where they feel most comfortable together. We call this spot ``$r_{eq}$," which stands for the distance where they're happiest.  Now, if they're too far apart, they start to feel a pull, like when you stretch a rubber band. That's why the energy goes up as they move away from each other. But if they get too close, they start to repel each other, like when two magnets push away. That's why the energy also goes up when they're too close. So, the curve shows us the sweet spot where these atoms are just right, not too close and not too far apart.}''  \\
\end{tabular}
\end{ruledtabular}
\end{table*}

In the above analogy, the student, while pointing at the graph, explains that at the equilibrium distance where the pink line meets the black line (albeit incorrectly), the spring is in its equilibrium position representing an ideal or a ``completely happy'' state in friendship between two friends. To the left of this point,  where the atoms seem to repel or the spring seeks to expand, the student compares it to a situation in which the friends spend excessive time together, until they get annoyed and start drifting away. To the right is where the atoms are attracted towards each other or where the spring tends to contract, the student compares it to friends who have been apart for some time and wish to reconnect. To the extreme end of the curve where the atoms overcome the electrostatic force, the student compares it to friends who have not seen each other in years, the bond of friendship may not be as strong as it used to be.

Through the lens of the Structure-Mapping theory, the above analogy has the {\em Target} domains of friendship and spring, with the {\em Base} domain as the curve. Atoms are mapped as friends, the distance between the atoms as the physical proximity between the friends, and the interactive forces as the bond of friendship. The higher order causal relational features are exhibited by the annoyance (or the repulsive force) leading to the drifting of friends and lack of contact (attractive force) leading to friends coming together.

\subsubsection{AI descriptions of the curve}

Unlike student-generated explanations, a predominant fraction of AI  descriptions employed comparisons, and all of these represented analogies. The predominant analogical contexts included magnets, friends, roller coaster, spring, tug-of-war, and slide.

Of the 839 AI-generated responses, 806 (96\%) contained comparisons, all of which reflected analogies. Among these, 190 included multiple embedded analogies within a single description. Only eight responses did not involve any comparisons, and twenty-five were excluded from analysis because their content was deemed contextually irrelevant, primarily due to improper prompting by students.  Figure~\ref{fig:q4-AI-pie-chart} summarizes these trends and Table~\ref{tab:AI-exemplar-responses} provides exemplar descriptions.

\begin{figure}
    \centering
   \includegraphics[scale=0.28]{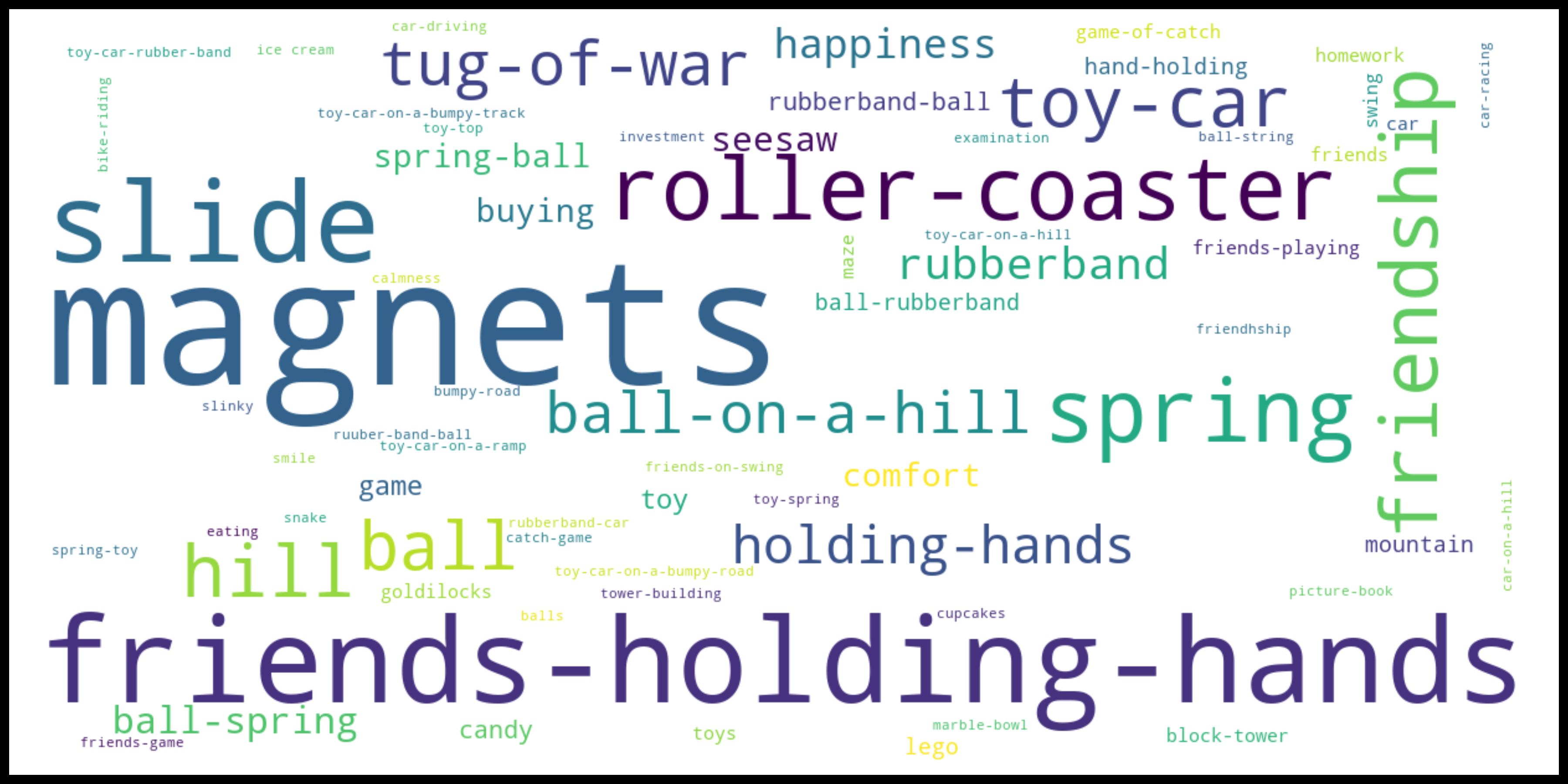}
    \caption{Word cloud representing the various analogical contexts employed highlighted by AI responses describing the Morse potential curve for second graders.}
    \label{fig:q4-AI-wordcloud}
\end{figure}

Figure~\ref{fig:q4-AI-wordcloud} presents the word cloud highlighting analogical contexts in AI descriptions. It can be observed that  magnets, friends, roller coaster, spring, tug-of-war, and slide were the predominant analogical contexts. Below is an exemplar AI-generated analogy of the curve in terms of two friends playing a tug-of-war game (along with a passing reference to a slide):

\begin{quote}
    ``{\em Imagine two friends, let's call them Hydro and Gen, who love playing a game of tug-of-war. They start by standing far apart, holding a rope between them. As they move closer, they feel more attracted to each other because they want to play together. This is like the potential energy going down on a graph as they get closer.  At some point, they find the perfect spot where they both feel happy and comfortable playing. This is like the equilibrium point on the graph—it's where they are in the best balance, not too far and not too close. It's like finding the perfect distance to stand while playing tug-of-war without falling over.  The lowest point on the graph is like the bottom of a slide where Hydro and Gen feel super happy and don't want to move. This is where they have the least energy and are most stable, just like when they're having the most fun playing together.  So, the graph helps us see how happy Hydro and Gen are at different distances, and when they're the happiest, they stay put and play together without any worries!}''
\end{quote}

The above description uses a tug-of-war game between two friends as the analogical context. The distance at which the friends feel comfortable while playing marks the equilibrium distance. This point is also similar to the lowest point on a slide, where the friends are content and prefer to remain there. As they move closer to each other, their desire to continue playing increases.

Through the lens of Structure-Mapping theory, the tug-of-war game presents the {\em Target} domain and the curve as the {\em Base} domain. The atoms are mapped as the two friends, the rope as the force, the desire to play as the potential energy, and the happiness between the friends as the stability of the atoms at the equilibrium point. Furthermore, the feature of the friends coming close to each other leading to their wish to play more underlines the higher order mapping of causal relationships.

\subsection{RQ2: Students’ curve descriptions in their preferred domains}
\label{subsec:q7-student}

We now unpack the trends in students’ self-generated analogies when explicitly prompted to describe the curve in  their preferred everyday contexts. Unlike students' descriptions of the curve to second graders, an overwhelming fraction of students employed analogies when explicitly prompted. Furthermore, the analogies were associated with diverse contexts ranging from sports to beekeeping. Among the descriptions which did not employ analogies, the preferred contexts involved disciplinary subjects such as chemistry and biochemistry. 

Of the 839 responses, 810 (96.5\%) generated comparisons with only 5 reflecting similarity statements and the rest 805 analogies. However, in only three of the 805 cases, students employed multiple analogical contexts to describe the curve. The remaining 29 descriptions did not entail analogies, as these curve descriptions involved the use of formal concepts with preferred contexts being highlighted as disciplinary domains such as chemistry, biochemistry, and engineering domains. Figure~\ref{fig:q7-pie-chart} summarizes these trends and Table~\ref{tab:q7-exemplar-responses} provides exemplar responses across each category.

\begin{figure}
    \centering
   \includegraphics[scale=0.28]{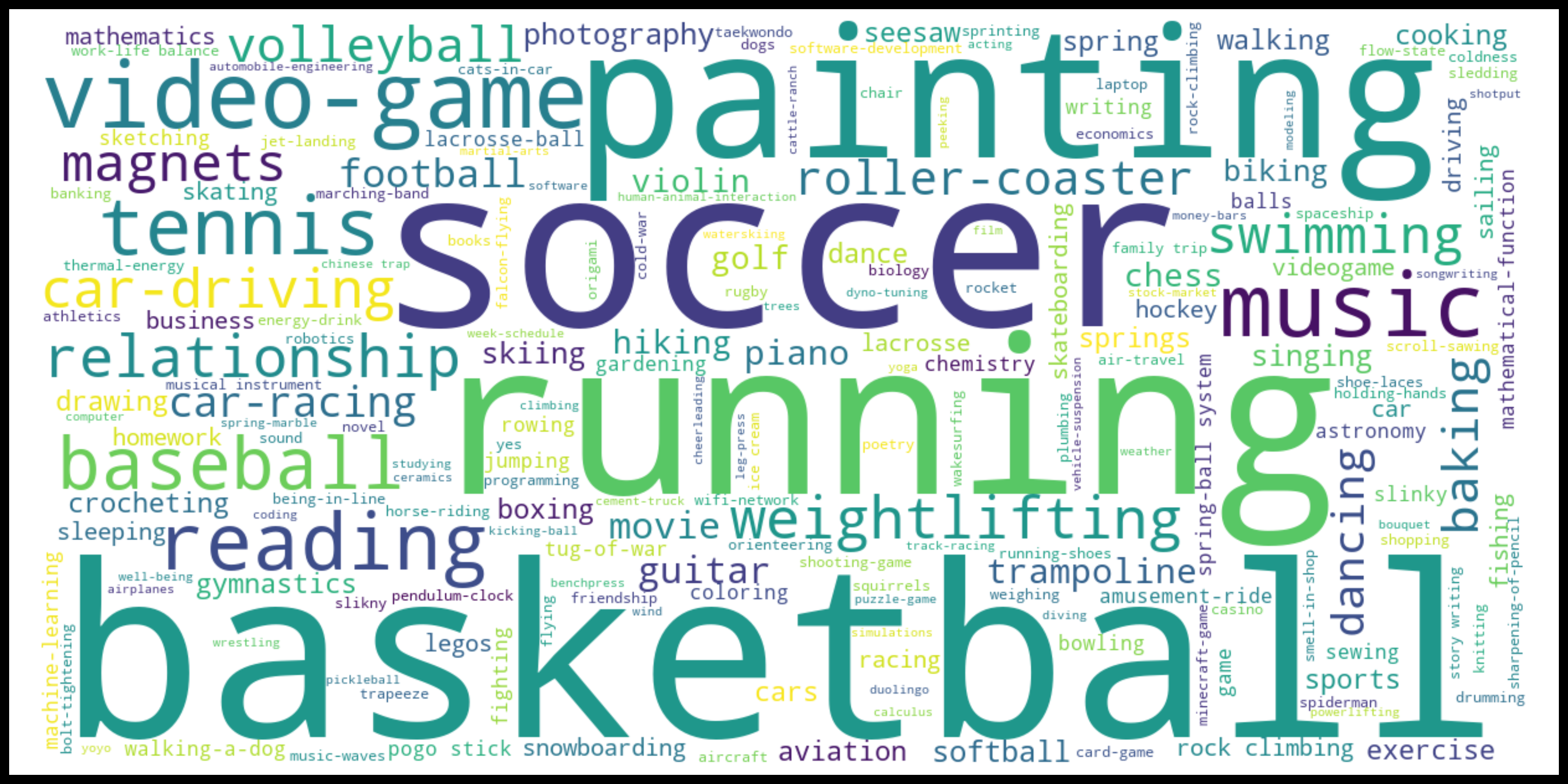}
    \caption{Word cloud representing the various analogical contexts employed by students when describing the Morse potential curve in terms of their preferred everyday contexts.}
    \label{fig:q7-wordcloud}
\end{figure}

\renewcommand{\arraystretch}{1.1}
 \begin{table*}[tb]
\begin{ruledtabular}
\caption{Exemplar student-generated responses highlighting explanations without comparisons and through similarity and analogy while explaining Morse potential curve in their preferred contexts.}\label{tab:q7-exemplar-responses}

\begin{tabular}{p{0.1\linewidth} p{0.87\linewidth}}

Category & Exemplar Quote  \\
\hline 

No comparisons & ``{\em In Chemistry, atoms and the stability of the bonds associated with them are often discussed, and this carries over to the practical applications of Chemistry as well. For example, there are ways to make carbon-based polymers more easily degradable by lengthening the carbon-carbon bonds in the polymer. This weakens the chain, making it more susceptible to decomposition. This can be visualized with the Morse Potential Diagram, where as r strays from $r = r_{eq}$, the energy of the bond is lowered, which means that it requires less energy to be broken. By this logic, we then find that $r = r_{eq}$ is the point in which the bonds are most stable.}'' \\

Similarity & ``{\em The domain of interest is a swimming pool, the req is the bottom of the pool.}''  \\

Analogy & ``{\em I am a beekeeper, so I will describe the graph in terms of bees. The curve represents how close a honeybee is to a flower. Initial the bee and the flower are very far apart, but the bee is attracted to the flower causing it to move closer to it. Once the bee lands on the flower, this is the smallest distance it will be from the flower, which is like the ``$r_{eq}$'' of the graph. Also, the bottom of the curve represents the smallest amount of effort the bee has to use to get to the flower. After the bee pollinates the flower, it leaves and goes away from the flower.}''  \\
\end{tabular}
\end{ruledtabular}
\end{table*}

Figure~\ref{fig:q7-wordcloud} presents the word cloud highlighting analogical contexts in student responses. Students employed a wide range of domains spanning sports (soccer, basketball, etc.), games (chess, video-game, etc.), music (instruments and singing), leisure (baking, reading, etc.), academics (software-development, robotics, etc.), relationships (e.g., family trip), and others (running shoes, air travel, etc.). 

The following is an exemplar student description of the curve in terms of reading a story:
\begin{quote}
    ``{\em My domain of interest is reading. To relate this to the diagram, we can think back to the reading flow charts we looked at in elementary school. The beginning of the story usually just gives details to the overall plot. This is like the start of the curve that is above the x axis. As you move along the curve, the plot starts to get interesting and the reader often has a lot of unanswered question. This is the area from the x axis to the peak of the curve. At the very top of the curve is the climax. In the diagram, this is when potential energy peaks. This would also be the peak of the story, or the turning point. As the curve approaches the x axis again, this would be the falling action of the story. Potential energy, or action in the story, decreases.}''
\end{quote}

In the above analogy, the student associates the process of reading a story with the shape of the graph. The start of the curve is compared to the beginning of the story where the details of the plot are introduced. As the story progresses, the plot gets interesting, which is represented by the bottom of the curve. The very peak (or bottom tip) symbolizes the story’s climax, where the potential energy peaks. As the graph moves along the x-axis, the ``action'' in the story decreases.

Through the lens of Structure-Mapping theory, the {\em Target} domain corresponds to reading and the curve as the {\em Base} domain. The curve is mapped onto the process of reading a story, potential energy with the interest of the reader, bottom tip of the curve with the story's climax, and the last section of the curve as the story's ending. The feature of building of the story to the climax leading to the increasing interest of the reader (increasing potential energy in the graph) marks the causal relationships reflected in the analogy.

\begin{figure}
    \centering
    \includegraphics[width=\linewidth]{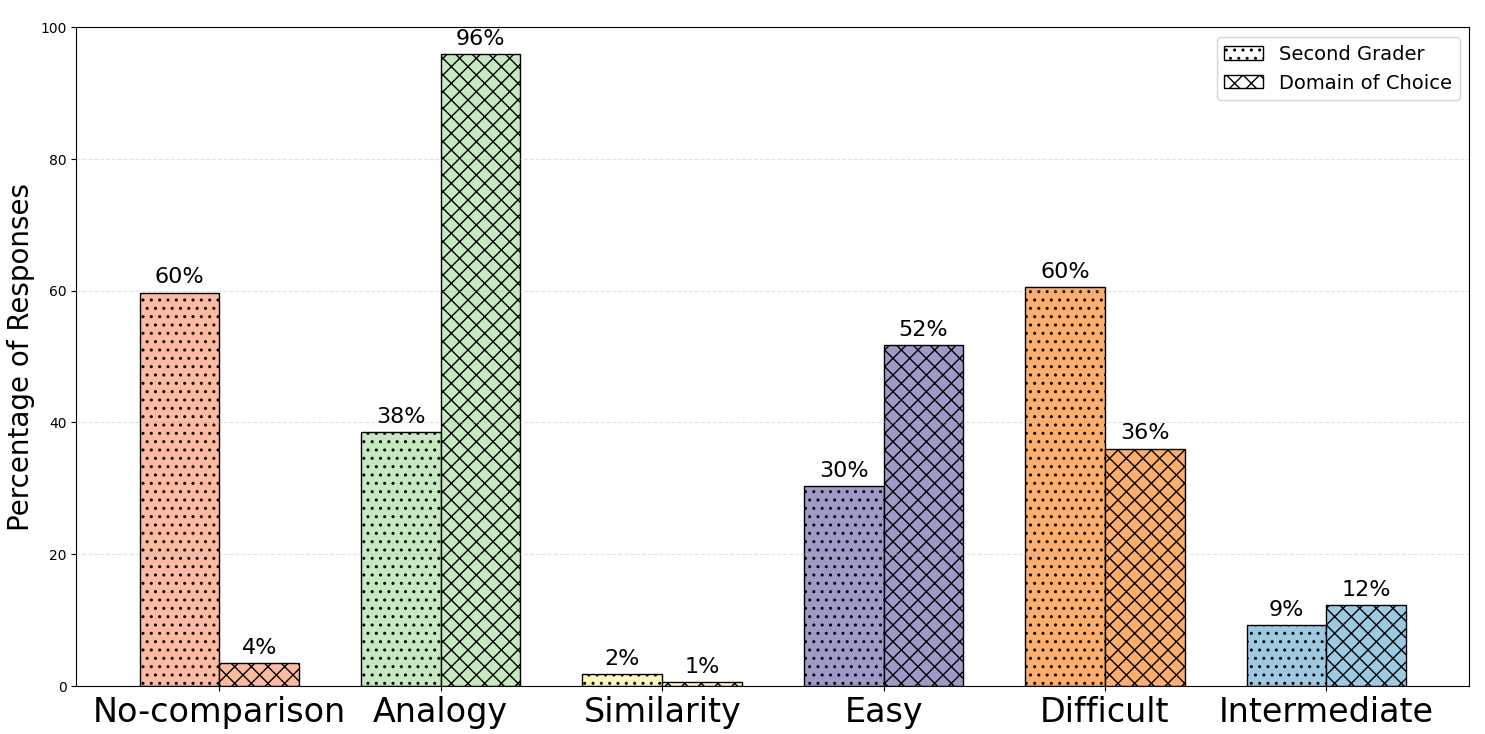}
    \caption{Comparative trends on students' use of comparisons and their perceived ease in explaining the curve in a language suitable for second graders and in their preferred contexts. }
    \label{fig:big-picture-comparison}
\end{figure}

\subsection{RQ3: Trends and themes in students’ perceived degree of ease in generating analogies}
\label{subsec:easy-difficult-themes}

We now turn to the third research question focused on trends and themes associated with students’ perceived ease and difficulties in explaining the curve to second graders and in their preferred contexts. As noted in Section~\ref{sec:methods}, students were asked about the perceived ease in generating explanations across the two contexts along with relevant reasons. We first begin with outline of quantitative distribution of observations followed by qualitative themes. Figure~\ref{fig:big-picture-comparison} highlights the quantitative trends across the two cases.

\begin{table}
\begin{ruledtabular}
\caption{Percentage of student responses highlighting perceived ease in generating explanations of the curve to second graders and in terms of their preferred contexts. }\label{tab:ease-comparison}
\centering
\renewcommand{\arraystretch}{1.2}
\begin{tabular}{clccc}

\multicolumn{2}{c}{} & \multicolumn{3}{c}{\textbf{Domain of choice}} \\ 

\multicolumn{2}{c}{}  & {Easy} & {Difficult} & {Intermediate} \\ 

\multirow{3}{*}{%
  \rotatebox[origin=c]{90}{\shortstack{\textbf{Second}\\\textbf{graders}}}%
} & Easy  & 15.6 & 10.8 & 3.8 \\

 & Difficult   & 31.3 & 22.2 &  7.0\\ 

  & Intermediate   & 4.8 & 3.0 & 1.4 \\ 
\end{tabular}
\end{ruledtabular}
\end{table}

\begin{figure}
    \centering
    \includegraphics[scale=0.35]{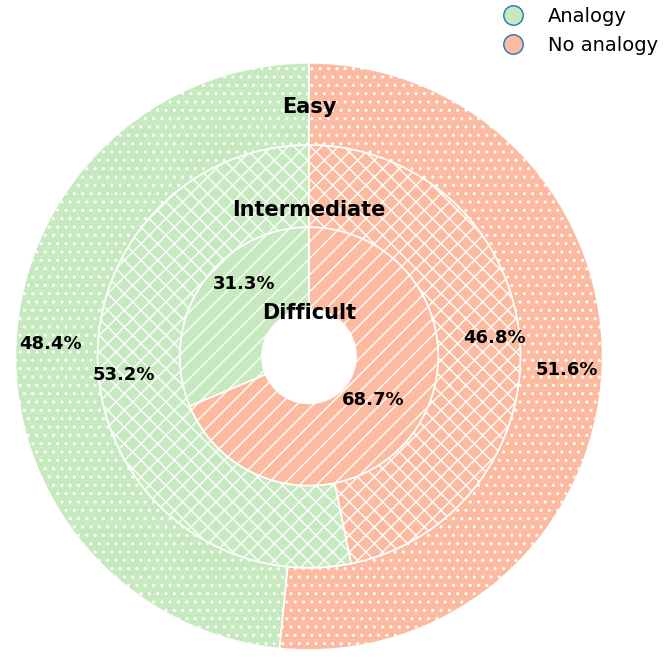}
    \caption{Comparative trends in themes associated with students' perceived ease and difficulty in explaining the Morse potential curve in a language suitable for second graders.}
    \label{fig:q4-perceived-ease-pie-chart}
\end{figure}

In the second grader context, around 30\% of the students perceived explaining the curve  as easy, 61\% as difficult, and 9\% as partly easy and difficult (``Intermediate''). Among the ones who perceived the process as easy or intermediate, only half generated analogies, whereas only one-third of the students who perceived the process as difficult did so. Figure~\ref{fig:q4-perceived-ease-pie-chart} summarizes these trends. These results, coupled with the observation of two-thirds of students not employing comparisons in their descriptions (Figure~\ref{fig:q4-piechart}), indicate that the process of explaining a complex idea to a non-disciplinary idea may not be an effective way of eliciting spontaneous analogies. 

\begin{figure}
    \centering
    \includegraphics[scale=0.35]{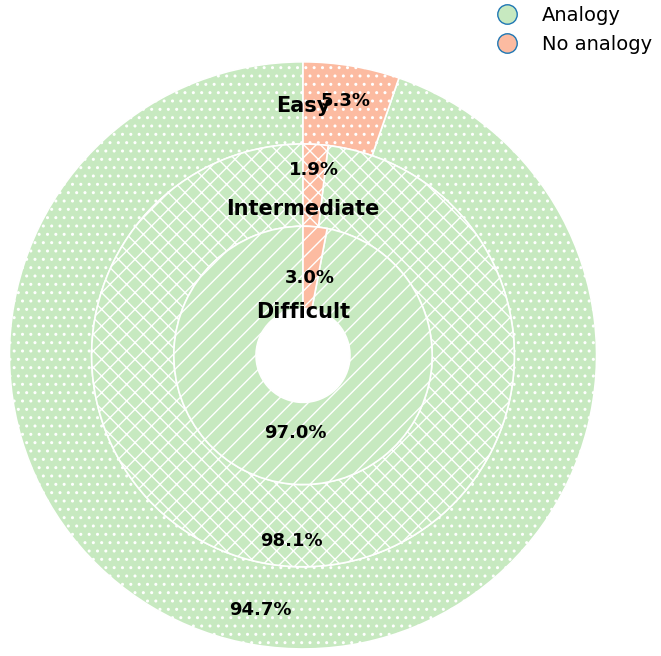}
    \caption{Comparative trends in themes associated with students' perceived ease and difficulty in explaining the Morse potential curve in their preferred everyday contexts.}
    \label{fig:q7-perceived-ease-pie-chart}
\end{figure}

In the case of explaining the curve in their preferred domains, the trend seemed to be reversed. While around 52\% perceived the process as easy, 36\% found it difficult, and 12\% described it as a mixture of both. Moreover, an overwhelming 95\%, 97\% and 98\% of the students who perceived the process as easy, difficult, and intermediate, generated analogies when explicitly prompted (Figure~\ref{fig:q7-perceived-ease-pie-chart}). These results, coupled with insights from the observations in Section~\ref{subsec:q7-student} regarding the diversity of analogical contexts, suggest that students exhibit a high degree of creativity and reasoning through varied analogies when explicitly encouraged to choose their own contexts.  

Table~\ref{tab:ease-comparison} highlights the comparative consistency among students on the perceived the ease of explaining the curve across the two cases. Around one-fourth perceived the process of generating explanations across both cases as difficult as opposed to mere 15\% perceiving as easy. In addition, one-third who perceived generating explanations to second graders as difficult, referred explaining in their preferred contexts as easy.  Only 15\% students perceived the process as easy in both cases.

Given these trends, we now discuss emergent qualitative themes in students’ responses about their degree of perceived ease in coming up with explanations across the two contexts. As noted above, we observed three broader categories across students’ descriptions about their perceived ease: ``Easy'', ``Difficult'', and ``Intermediate''. Below we begin with themes associated with these categories across the two cases. 

\subsubsection{Second graders}

We find three themes associated with the ``Easy''  and four each across ``Difficult'' and ``Intermediate'' categories associated with students’  perceived ease in generating explanations about the curve to second graders. Below, we briefly describe these themes. Table~\ref{tab:q4-themes} provides exemplar quotes that evidence each of the below described themes.

Among the students who perceived the process as “Easy” (27\%),  a predominant subset attributed this to their strategy of simplifying terminology (i.e., ``dumbing down the vocabulary’’) or focusing only on selected elements of the graph. This theme aligns with observations from Figure~\ref{fig:q4-piechart}, which shows that 60\% of students did not employ analogies when describing the curve to second graders. The second theme related to the degree of conceptual understanding that shaped students’ perceptions of ease. Many students found the task easy either because they had a strong understanding of the concept or because they could articulate explanations that reflected the limited extent of their understanding. The last theme corresponded to students' personal attributes or experiences towards generating explanations. These mainly corresponded to their experiences interacting with second graders (with siblings or as elementary teachers) or their own personal learning approaches of frequently coming up with analogies for understanding complex concepts.

\renewcommand{\arraystretch}{1.5}
 \begin{table*}
\begin{ruledtabular}
\caption{Themes and exemplar quotes associated with students' perceived ease in explaining Morse potential curve in a language suitable for second graders.}\label{tab:q4-themes}

\begin{tabular}{p{0.04\linewidth} p{0.35\linewidth} p{0.55\linewidth}}

Theme & Description & Exemplar quote \\
\hline 

\textbf{Easy}    & \textbf{}  & \\

E1    & Frequently employs analogies as a personal strategy to think, explain, or understand concepts.  & ``{\em I found it kind of easy to describe the curve to a second grader. I am used to being around kids and lot so I think it comes easy to me thinking about things in a way that would help them understand.} [...]'' \\

E2    & Simplifies terminology or focuses on key elements to make ideas easier to grasp. & ``{\em I found it a little bit easier because I am able to dumb down what the graph actually is. Also, it helped me to understand the graph, I guess because prior to this, I didn't know the graph all that well.}'' \\

E3    &  Given concept is easy to understand, describe, or intuitive. & ``{\em I found it easier because the energy principle is something that can be described in simple ways. Since energy cannot be created or destroyed, the concept of explaining it goes as far as saying that if one part goes up, the other goes down.}[...]'' \\

\textbf{Difficult} & \textbf{} & \\ 

D1      &  Issue with simplifying vocabulary or terminology with or without losing conceptual accuracy. & ``{\em It's a bit more difficult. Yo have to be able to explain it in term that a second grader would know without over simplify or leaving out certain information.}'' \\

D2     & Lack of a thorough understanding of the curve OR Concern that simplifying the concept could create confusion or hinder their own understanding. & ``{\em I found it very difficult, surprisingly not because of the complex language, but more so because of the fact that I really don't have a good enough grasp of the subject to be able to explain it.}[...]'''  \\

D3     & Issue with explaining, particularly abstract and non-tangible ideas. & ``{\em I found it more challenging because I am not sure how to explain the concept of potential energy in terms a second grader may understand. This is most likely because potential is a sort of obscure concept to visualize.}''\\

D4     & Unsure about the level of understanding of second graders OR Complex concept to explain, given the age of the target audience OR Lack of personal familiarity with second graders' understanding. & ``{\em I found it difficult to describe the curve to a second grader. The first challenge I experienced was trying to figure out what their knowledge basis was. It was difficult for me to conceptualize their frame of reference and explain in a manner that I feel was effective.}[...]'' \\

\textbf{Intermediate} & \textbf{} & \\ 

I1      &  Certain terms or elements of the concept are easier to explain than others OR Well-understood ideas feel easy to explain, the unclear ones feel difficult. & ``{\em It was a little hard to explain potential and kinetic energies, but the actual curve is easy to explain once that is out of the way.} [...]   \\

I2     & Initially difficult, but becomes easier once ideas are simplified or an analogy is created. & ``{\em I found it difficult at first because I am used to things being explained at a college level now. But, once I sat and thought about it for a minute, I was able to think of a scenario to represent the attraction and repulsion between two atoms. I thought of this scenario because I have siblings and have experienced missing them sometimes, but also wanting to be away from them sometimes.}'' \\

I3     & A combination of existing easy and difficult codes. & ``{\em Yes and no. It made sense in my mind having to break it down but at the same time it loses a lot of important information. Vital aspects to the graph are lost simply due to the missing context. This is the challenging portion.}'' \\

I4     &  Easy because of time availability, else difficult. & ``{\em I found it slightly easier because I had more time to think about the previous question but if I went straight to describing the req to the second grader then it would've been much harder. It is definitely slightly harder to describe physics to a second grader because I know a lot of the physics concepts to understand the graph, but it's hard to simplify all of it so that someone that young can understand it.}'' \\
\end{tabular}
\end{ruledtabular}
\end{table*}

\renewcommand{\arraystretch}{1.5}
 \begin{table*}
\begin{ruledtabular}
\caption{Themes and exemplar quotes associated with students' perceived ease in explaining Morse potential curve in terms of their preferred contexts.}\label{tab:q7-themes}

\begin{tabular}{p{0.04\linewidth} p{0.35\linewidth} p{0.55\linewidth}}

Theme & Description & Exemplar quote \\
\hline 

\textbf{Easy}    & \textbf{}  & \\

E1    & Overlap between the context and concepts OR The curve is easy to visualize or intuitive in the context.  & ``{\em I found it much easier to describe the terms in my domain of interest because I was able to choose an analogy for which I could easily locate req and connect the quantities of kinetic and potential energy (through pages of a book). }'' \\

E2    & Familiarity of concepts through everyday experiences, background formal knowledge, provided information, or broadly applicable ideas.    & ``{\em I found it easy in terms of basketball because I feel that this curve can apply to many things in life. There is a balance in every aspect of life and when crossed will lead to more negative outcomes than positive. This ultimately causes you to back off and revert back to the optimal balance.} [...]'' \\

E3    & Flexibility to adapt analogies based on preferred contexts, understanding, assumptions, or vocabulary.   & ``{\em I found it easy. It was a different way of thinking about physics and I enjoyed relating it to real life and something I am passionate about. It forced my brain to make connections and get a deeper understanding when trying to explain it to someone else using a metaphor.}'' \\

E4    & Ease of mapping facilitated by prior knowledge, experience, tangible representations, or contextual supports.   & ``{\em I found it easy because after describing the curve multiple times, the idea gets much simpler and you can relate it to almost anything that has a balance between too far and close, like a seesaw.}'' \\

E5    & Self-perception as creative/intuitive or experience with connecting physics to everyday experiences.   & ``{\em I feel it is easier to describe the curve in an interest of my choice because that is how my mind works to understand things. When faced with something confusing, I try to break it down or rethink it in terms of things that I do understand, so describing this in a way that I get was basically doing just that.}'' \\

{\bf Difficult}  & \textbf{}  & \\

D1    & Challenges in aligning physics concepts with the chosen context or communicating ideas effectively.  & ``{\em Hard, very hard. Explaining physics ideas without using physics terms is extremely hard. When trying to explain any area of STEM in an artistic manor is difficult for understanding these physical properties requires more than feelings and thoughts. It require background knowledge and practice.  }'' \\

D2    & Difficulty identifying or constructing suitable analogies and real-world connections.  & ``{\em It was very difficult, but I described in something that interests me, so it made it bearable. It was hard to come up with analogies for this type of thing, because it's hard to explain other than just stating what it is. }'' \\

D3    & Uncertainty or lack of conceptual clarity about the task, curve, or the generated analogy.  & ``{\em I found it very hard partially because I am having difficulty describing the curve before putting it in terms of my domain of interest and you must really understand the curve before you can explain it in non-physics terms.}'' \\

D4    & Limited confidence or experience in generating analogies.  & ``{\em Very very difficult. I am not very creative with analogies.}'' \\

{\bf Intermediate}  & \textbf{}  & \\

I1    & Difficult to come up with analogical contexts or mapping, but easy to explain.  & ``{\em Not really, it did take minute to come up with an idea to explain what I wanted to say but after I got my idea, the analogy flowed. }'' \\

I2    & Well-understood ideas are easier to map, others are difficult OR Easier to map few but difficult to map all.  & ``{\em I found the part that I think I understand well to be extremely easy, but the part I feel pretty shaky (and pretty wrong) about I found to be very difficult. }'' \\

I3    &  Combination of easy and difficult codes. & ``{\em I found it difficult to understand what the question was asking me to do at first, but once I figured it out I found it easy and rather enjoyable because I related it to something I enjoy.}'' \\
\end{tabular}
\end{ruledtabular}
\end{table*}

Among the ones who perceived the process as ``Difficult'', a predominant section focused on the barriers associated with understanding their target audience, i.e., second graders. Many attributed their difficulty to a lack of familiarity with the cognitive level of second graders or limited experience interacting with that age group. The second major theme related to challenges in simplifying terminology without compromising conceptual accuracy. Students expressed difficulty in striking a balance between conceptual rigor and using age-appropriate language. The third theme reflected students’ limited conceptual understanding of the curve. These students noted that attempts to simplify the concept often led to confusion due to their own incomplete understanding of the concept. The final theme highlighted individual attributes, with several students describing themselves as being ``bad'' at explaining ideas, particularly abstract and non-tangible ones.

For the last category, i.e., those who perceived the process as ``Intermediate'',  the first theme pertained to students’ perceived ability to explain only a subset of ideas effectively. For example, they found it easier to explain concepts they understood well and more challenging to articulate those they found unfamiliar. The second theme centered on the sequential nature of the explanation process: students reported that while it was initially difficult, it became easier once ideas were simplified or analogies were constructed. The third theme was marked by a blend of easy and difficult themes described in the previous two categories. The last theme corresponded to students attributing time factor as crucial. These students (though very less in number) highlighted that if they had enough time, the process would be relatively easy, else difficult. 

\subsubsection{Domain of Choice}

In the case of students’ perceived ease  of explaining the curve in their preferred contexts, we identified five themes associated with the ``Easy'', and four themes each for ``Difficult'' and ``Intermediate'' categories. Table~\ref{tab:q7-themes} provides exemplar quotes that evidence each of the themes that are detailed below. 

Among the ones who highlighted the process as ``Easy'', a predominant section attributed it to the perceived agency in choosing the context and explaining the curve using their own vocabulary. Students in this theme highlighted the freedom to pick their preferred contexts, use relevant vocabulary (either of physics or of their contexts), make broad assumptions, and selectively focus on certain concepts by overlooking others. The second major theme corresponded to the extent of overlap between the context and features of the curve. Students highlighted that their preferred context shared several features of the curve, enabling them to better visualize it in their chosen contexts. 

The third theme corresponded to the cluster of factors that facilitated students in generating analogies. These focused on their understanding about physics as well as the context, the presence of tangible physics ideas enabling them to better explain, the feature of repeatedly explaining the curve first in their own words and then to second graders in the exercise, and the curve being taught as an analogy in the course curriculum. The penultimate theme corresponded to the features of the curve that facilitated the generation of analogies. These included: the  involved concepts being extensively observed in daily life, the concepts being generic enough to map onto various domains, or the provided image of the curve being detailed enough to help generate analogies. The final theme reflected personal attributes or practices towards generating analogies such as perceiving themselves as creative, instinct guiding selection and explanation of analogy, or their prior experience generating analogies.

Among the students who perceived the process as ``Difficult'', we observed four distinct themes. The first major theme corresponded to the difficulty in mapping between the features of the curve with the elements of their chosen contexts. These included: non-alignment or partial alignment between the curve and the context, using the right vocabulary to communicate the curve to the general audience, and trouble mapping two unrelated ideas. The second theme corresponded to coming up with the analogy in the first place. Students highlighted that it was difficult to identify, visualize, or justify the chosen context, or difficult to come up with real-world analogs of ideal models such as the curve. The third theme highlighted other distinct difficulties in generating customized analogies such as the lack of content knowledge about the curve, lack of understanding about the question, lack of clarity about the correctness of the analogy, and not being able to use visuals to better communicate ideas. The last theme reflected personal attributes such as students perceiving themselves as not better at generating analogies or their lack of experience with coming up with analogies. 

The third category, i.e., ``Intermediate'' (a combination of easy and difficult) entailed four major themes. The first predominant theme focused on students perceiving difficulty in selecting appropriate analogical contexts but the remaining explanation process as easy. The second theme corresponded to the perceived ease in mapping selective ideas such as the ones well-understood as compared to others. The third theme corresponded to a blend of codes from the previous ``Easy'' and ``Difficult'' themes described above. The last theme corresponded to students’ perceptions of the explanation generation process as being easy but however not being confident of the correctness of the resultant analogies.

\section{Discussion}
\label{sec:discussion}

We qualitatively analyzed how student- and AI-generated responses (prompted by students) explained Morse potential curve in a language suitable for second graders. We hypothesized that students would ``spontaneously’’ employ analogies while explaining such a complex idea to second graders. The objective to seek similar responses from AI was to explore the features of AI-generated explanations as compared to student-generated ones, and thus analyze its potential to support students in generating spontaneous analogies. We also analyzed students’ explanations of the curve in their preferred everyday contexts, with the underlying objective to their ``self-generated’’ analogies. Lastly, we examined themes associated with students’ perceived ease in generating explanations across the two scenarios. 

Contrary to our hypothesis, results highlight that only around one-third of the students constructed spontaneous analogies while explaining the curve to second graders. Among the generated analogies, the predominant context involved springs, the same context which was also illustrated in the curve provided during the activity (Figure~\ref{fig:morse-curve}). The remaining two-thirds explained the curve by either simplifying the formal vocabulary or by generating elaborate explanations. Students’ reflections on the perceived ease also highlighted the same, as a common theme across the ``Easy’’ and ``Difficult’’ categories corresponded to translating formal vocabulary into simpler words. 

The AI responses on the other hand, supported our hypothesis with overwhelming fraction of responses (around 96\%) entailing analogies when prompted by students to explain the curve to a second grader. In addition, a larger fraction (around one-fourth) of the AI-generated analogies entailed multiple analogical contexts embedded in single explanations as compared to students (around one-tenth). These observations highlight that nudging students to explain a complex idea to a non-disciplinary audience may not be an effective approach to elicit spontaneous analogies, and AI presents a strong potential to assist students in this objective.

In contrast, when explicitly asked to explain the curve in their preferred everyday contexts, an overwhelming fraction (around 96\% of 839 students) came up with self-generated analogies, that too involving diverse contexts. This result was observed despite around half of the students perceiving the explanation generation as difficult. Though mapping formal ideas onto their chosen contexts was highlighted as a major barrier, students also noted the agency in choosing their own contexts and explaining in their own vocabulary as a predominant factor making the process easier. These observations, particularly students generating diverse analogies despite perceiving it as difficult, highlight the stark contrast between students' spontaneous and self-generated analogical reasoning. 

These results contribute to the contemporary literature on student-generated explanations, particularly spontaneous and self-generated analogies. Pitterson {\em et al.}~\cite{pitterson2018engineering} observed that when engineering students spontaneously generated analogies, they mainly drew the contexts taught in their introductory classes. Sandifer~\cite{sandifer2004spontaneous} observed incorrect understanding of the analogy’s {\em Base} domain as interfering with students’ generation of spontaneous analogies in physics. Furthermore, several studies in cognitive science too highlight that despite the presence of useful relational matches in long term memory, humans typically struggle to retrieve them~\cite{gick1980analogical,gick1983schema}. Our results overlap with these studies as a large fraction of students in our data too struggled to generate spontaneous analogies and the ones generated were predominantly guided by contextual cues. Our results also align with Sandifer’s observation as the lack of content understanding of the curve was one of the themes associated with students’ perceived difficulty in explaining the curve to second graders.

In addition to aligning with existing observations, our findings shed additional insights on perceived difficulties experienced by students across the three phases of analogical retrieval (Section~\ref{subsec:analogical-retrieval}), especially while generating spontaneous analogies. As for the first ``{\em Retrieval}” phase (retrieving appropriate analogical context from long term memory), students highlighted the issue of analogizing abstract and non-tangible ideas in physics. In the ``{\em Mapping}'' phase (establishing associations between base and target domains), students highlighted simplifying vocabulary at the cost of conceptual accuracy or their lack of content understanding about the curve as the perceived barriers. Lastly, for the ``{\em Evaluation}’’ phase (evaluating and validating the relevance of generated analogy), students highlighted concerns about the validity of the generated explanation given the content understanding of the target audience, i.e., second graders.  Additionally, our results also highlight that students’ perceived ease is not always binary (i.e., ``easy’’ or ``difficult’’) but can also represent the blend of both where some features tend to facilitate analogical reasoning than others.  We also observed individual or personal beliefs also influencing their perceived ease in generating spontaneous analogies.

Our results also contribute to the growing body of literature focused on exploring AI's potential for facilitating students’ analogical reasoning. Cao {\em et al.}~\cite{cao2023elucidating} explored the potential of AI platforms in elucidating STEM concepts through analogies in natural language as well as through AI-generated analogical images. The authors observed that while AI platforms were better at generating analogies in natural language, generating corresponding image representations presented a challenge. Though our study did not explore AI generated analogical images, our results overlap with Cao {\em et al}’s observations on the feasibility of AI in generating both spontaneous and self-generated analogies, though conceptual accuracy need to be further explored. 

Bernstein {\em et al.}~\cite{bernstein2024like} and Shao {\em et al.}~\cite{shao2025unlocking} further explored the potential of AI in generating analogies involving contexts specified by students. Bernstein {\em et al.}~\cite{bernstein2024like} observed first-year computing students specifying a diversity of analogical contexts to ChatGPT while trying to generate recursion-based analogies by building on a provided code snippet. Shao {\em et al.}~\cite{shao2025unlocking} explored the feasibility of generating analogies using AI about high-school level physics and biology concepts. The authors observed that compared to biology, AI-generated analogies about physics content tend to be conceptually incorrect. Similar to observations from Bernstein {\em et al.}, students in our data too leveraged diverse everyday contexts while generating analogies. However, unlike observations from Shao {\em et al.}, preliminary analysis of the AI-generated analogies in our context seemed to be relatively conceptually correct. Validating this claim however warrants additional work. 

\section{Conclusions, Limitations, and Future Work}
\label{sec:conclusions}
Analogies are central to practice and communication of physics as they aid in integrating new information by building on one's existing ideas. In classroom practices, self-generated analogies are more conducive for students’ learning than the ones explicitly taught. ``Spontaneous’’ and ``self-generated’’ analogies represent the two ways through which students generate their own analogies. In this study, we explore the extent to which students spontaneously generate analogies when asked to explain a complex idea such as Morse potential curve in a language suitable for second graders. We compare student explanations with AI responses towards the same objective. We also explore how students explain the curve when explicitly asked to describe it in terms of their preferred everyday contexts. Lastly, we qualitatively analyze the themes associated with students’ perceived ease and difficulties in generating explanations across the two cases.  Results highlight that unlike AI responses, student-generated explanations to second graders seldom employ analogies. However,  when explicitly asked to describe the curve in their everyday domains, students generate analogies by employing diverse contexts. A combination of knowledge and personal attributes tend to influence students’ perceived ease in generating explanations across the two cases. 

Findings reported in this study present several implications for teaching and learning of physics. Firstly, the potential of AI in facilitating students’ generation of personalized analogies. Given the recent and rapid development of AI capabilities, educators have called for leveraging this technology for promoting the objectives traditionally valued by the learning sciences community~\cite{walter2024embracing}. Analogical reasoning, particularly students’ self-generated analogies represent one such objective. Our results highlight that AI models can generate simplified explanations through embedded analogies describing complex ideas in simple words. Activities focused on comparing and critiquing self- and AI-generated explanations, particularly in a language suitable for non-disciplinary audience presents a promising way of effectively leveraging AI for college physics learning. Such exercises can also facilitate students in engaging science outreach activities or creating informal physics learning environments. 

Secondly, as highlighted by the themes on students’ perceived ease in generating explanations, such exercises can make students notice gaps in their knowledge about their content understanding of the given concepts. Along with ascertaining this phenomenon across multiple learning contexts, researchers can also unpack the cognitive processes guiding sensemaking through analogical reasoning in physics.  Furthermore, we observe that a large fraction of students perceived the process of explaining complex ideas in simpler words as ``difficult’’, thus highlighting the potential of such exercises in promoting ``productive struggles’’ in physics classroom learning. Such exercises can be part of formative assessments to capture content understanding of students, particularly of complex concepts. 

Lastly, we observed students extensively employing diverse analogical contexts while communicating physics ideas. Collaborative classroom activities focused on students with similar interests individually generating their analogies and then providing constructive critique to their peer’s analogies presents potential for students’ active engagement and participation in classrooms. For environments with smaller classroom sizes, AI models can play the role of ``learning partners’’ in generating analogies on students' preferred contexts (such as sports, music, leisure, etc.) and students critiquing them. We also observe students’ attributing personal experiences of interacting with second graders as influencing their perceived ease in generating explanations. Such activities can thus be an integral component of pre-service teachers’ education, targeting both pedagogical skill of explaining through analogies while simultaneously attending to their content understanding in physics. 

Our findings accompany several limitations. To begin with, many students highlighted that they were unsure about referring learning materials while responding to the tasks. Given that the activity was online and students received credits for active participation, we assumed and expected students to refer learning materials but did not explicitly clarify the same. Student responses generated through varying use of learning materials during the activity may have influenced the reported results. Secondly, the sequence of the activity. As described in Section~\ref{sec:methods}, students first described the curve in their own words before describing the same to second graders and then in terms of their preferred domains. Students’ perceived ease in explaining the curve in their own contexts may be influenced by the prior activity of explaining to second graders. Lastly, analogical reasoning is a complex activity extending beyond language use, such as communicating through visuals or other representations. We acknowledge that students have varying learning preferences (also explicitly acknowledged by students in our data about their preference to convey the curve through visual aids rather than words), and results reported in this study are constrained by the medium of students’ expression.

Future work would focus on addressing the above limitations along with expanding our work to analyze remaining parts of the activity. As highlighted in Figure~\ref{fig:data}, the activity also involved students comparing their and AI-generated responses of the curve, description to second graders, and in terms of their preferred contexts. Students also highlighted the perceived usefulness of the activity for their own content understanding of the curve. Since we also collected data on students’ prompts to AI, we seek to explore how the quality of the prompts influenced AI responses and by consequence, students’ learning during the activity. We also plan to follow up with studies focused on leveraging unsupervised machine learning approaches (such as K-means clustering) and large language models in capturing additional insights in our data. Such efforts present strong potential to effectively promote valued objectives such as analogical reasoning in the AI era.

\section{Acknowledgments}
This study was supported by Purdue University's Innovation Hub and U.S. National Science Foundation grant 2111138. Any opinions or findings expressed here are those of the authors and not of the Foundation.

%

\end{document}